\documentclass[prl, aps, nofootinbib, twocolumn, amssymb, superscriptaddress, 10pt]{revtex4-2}
\usepackage{amsmath}
\usepackage{amssymb}
\usepackage{amsthm}
\usepackage{amsfonts}
\usepackage{enumerate}
\usepackage{latexsym}
\usepackage[dvipsnames]{xcolor}
\usepackage{setspace} 
\usepackage{blindtext}
\usepackage{dsfont}
\usepackage{mathrsfs}
\usepackage[normalem]{ulem}

\usepackage{physics}
\usepackage{tabularx}
\newcolumntype{Y}{>{\centering\arraybackslash}X}

\usepackage{stackengine}
\usepackage{bbold}

\usepackage{bm}
\usepackage{graphicx}

\usepackage{hyperref}
\hypersetup{
pdfnewwindow=true, colorlinks=true,
linkcolor=red, anchorcolor=res,
citecolor=red, filecolor=red,
menucolor=red, urlcolor=red} 

\usepackage{pifont}

\newcommand{\beginsupplement}{
        \setcounter{table}{0}
        \renewcommand{\thetable}{S\arabic{table}}
        \setcounter{figure}{0}
        \renewcommand{\thefigure}{S\arabic{figure}}
        \setcounter{equation}{0}
        \renewcommand{\theequation}{S\arabic{equation}}
        \setcounter{section}{0}
        \renewcommand{\thesection}{\Alph{section}}
        \setcounter{subsection}{0}
        \renewcommand{\thesubsection}{\arabic{subsection}}
}

\newcommand{\vk}{{\mathbf{k}}}

\newcommand{\qs}[1]{{\color{red} #1}}

\begin{document}

\title{Correlated flat-band physics in a bilayer kagome metal based on compact molecular orbitals}

\author{Mounica Mahankali}
\thanks{These two authors contributed equally}
\affiliation{Department of Physics and Astronomy,  Extreme Quantum Materials Alliance, Smalley Curl Institute, Rice University, Houston, Texas 77005, USA}
\author{Fang Xie}
\thanks{These two authors contributed equally}
\affiliation{Department of Physics and Astronomy,  Extreme Quantum Materials Alliance, Smalley Curl Institute, Rice University, Houston, Texas 77005, USA}
\author{Yuan Fang}
\affiliation{Department of Physics and Astronomy,  Extreme Quantum Materials Alliance, Smalley Curl Institute, Rice University, Houston, Texas 77005, USA}
\author{Lei Chen}
\affiliation{Department of Physics and Astronomy,  Extreme Quantum Materials Alliance, Smalley Curl Institute, Rice University, Houston, Texas 77005, USA}
\author{Shouvik Sur}
\affiliation{Department of Physics and Astronomy,  Extreme Quantum Materials Alliance, Smalley Curl Institute, Rice University, Houston, Texas 77005, USA}

\author{Silke Paschen}
\affiliation{Department of Physics and Astronomy,  Extreme Quantum Materials Alliance, Smalley Curl Institute, Rice University, Houston, Texas 77005, USA}
\affiliation{Institute of Solid State Physics, Vienna University of Technology, Wiedner Hauptstr. 8-10, 1040, Vienna, Austria}

\author{Jean C. Souza}
\affiliation{Department of Condensed Matter Physics, Weizmann Institute of Science, Rehovot, 7610001, Israel}

\author{Moshe Haim}
\affiliation{Department of Condensed Matter Physics, Weizmann Institute of Science, Rehovot, 7610001, Israel}

\author{Ambikesh Gupta}
\affiliation{Department of Condensed Matter Physics, Weizmann Institute of Science, Rehovot, 7610001, Israel}

\author{Nurit Avraham}
\affiliation{Department of Condensed Matter Physics, Weizmann Institute of Science, Rehovot, 7610001, Israel}

\author{Haim Beidenkopf}
\affiliation{Department of Condensed Matter Physics, Weizmann Institute of Science, Rehovot, 7610001, Israel}

\author{Hengxin Tan}
\affiliation{Department of Condensed Matter Physics, Weizmann Institute of Science, Rehovot, 7610001, Israel}

\author{Binghai Yan}
\affiliation{Department of Condensed Matter Physics, Weizmann Institute of Science, Rehovot, 7610001, Israel}

\author{Qimiao Si}
\affiliation{Department of Physics and Astronomy,  Extreme Quantum Materials Alliance, Smalley Curl Institute, Rice University, Houston, Texas 77005, USA}

\date{\today}

\begin{abstract}
Flat bands, when located close to the Fermi energy, can considerably enhance the
influence of electron correlations on the low energy physics in kagome and other frustrated-lattice metals. 
A major challenge in describing the interaction effects in such bulk materials is that the flat band is often intermixed with a large number of other bands.
Here we show that the recently introduced notion of compact molecular orbitals (CMOs) enable a path forward in describing the dominant 
effect of the Coulomb interactions in spite of the complexity 
of the bandstructure. 
Our materials-based analysis allows for the understanding of
the scanning-tunneling-microscopy 
experiment \citep[J. C. Souza et al., preprint (2024)]{exp-stm-2024} 
of the bilayer kagome metal Ni$_3$In in terms of the CMO notion.
From the resulting CMO,
an effective Anderson lattice model can be set up.
This CMO-based approach enables the calculation of correlation effects
that is difficult to do based on the atomic orbitals.
Furthermore, it suggests an enriched phase diagram for the strange metal physics 
of the kagome metal, which can be tested by future experiments.
We discuss the implications  of our results for the general correlation physics 
of flat band systems and beyond.
\end{abstract}

\maketitle

{\it Introduction.~~~}
Strong correlations develop in quantum materials when the Coulomb interaction
reaches or exceeds the width of electron bands \cite{Keimer2017,paschen_quantum_2021}.
This happens in transition metal compounds such as the cuprates, which is a  local-moment-bearing
Mott insulator at half filling and a strange metal under carrier doping for optimal superconductivity \cite{Lee-RMP06,Phillips2022Strange}.
Heavy fermion metals represent another prototype, containing multiple atomic orbitals
that are active in the low energy physics. They feature a very narrow $f$-electron band
and wide $spd$ bands.  When the Coulomb interaction is larger than the $f$-bandwidth
but smaller than the width of the $spd$ bands, a physically transparent model arises, 
namely the Kondo lattice model of $f$-orbital local moments coupled to $spd$-orbital itinerant electrons \cite{hewson1997kondo}.
The model features strange metallicity through the beyond-Landau quantum criticality
of Kondo destruction  \cite{Hu-Natphys2024,Si2001,Colemanetal,senthil2004a}.
This description  of strange metallicity allows for a unified understanding of
a variety of measurements
concerning such features as dynamical scaling, Fermi surface reconstruction, and
loss of quasiparticles \cite{Kirchner_RMP,Si10.2,Aro95.1,Schroder00,Prochaska2020,paschen2004,Friedemann09,shishido2005,Pfau2012,Chen2023Shot}.
This paradigm of Kondo destruction also connects with the orbital-selective Mott physics, which has been extensively discussed in systems such as 
Fe-based superconductors \cite{Si2023Iron}.

Recently, for transition-metal-based systems on kagome and related frustrated lattices, 
the framework of local-moments coupled to itinerant electrons has been developed in terms of compact molecular 
orbitals (CMOs)
\cite{Chen2023Metallic,Chen2024Emergent,Hu2023Coupled}. Here, a CMO represents a linear superposition of atomic orbitals from
different sites of a lattice
such that it is symmetry preserving.
A CMO captures the destructive interference that underlies a flat band, 
as the intuitive notion of compact localized state \cite{Leykam_local-mode2018,Bergman2008} does.
In contrast to the latter, the CMOs are orthonormal and, together with its more extended counterparts, form a complete basis in real space to represent the combined flat and wide bands. In comparison to molecular orbitals that arise in other contexts, such as for Kitaev systems \cite{Mazin_Kitaev2012}, organic conductors \cite{Ferber_organic2014}
and Mott insulators \cite{Grytsiuk_Mott2024}, 
the CMOs emphasize on the flat bands and their coupling 
with wide bands.
Out of this interplay,
local moments develop from the CMOs and, in combination with the more itinerant electrons 
of the molecular orbitals that are orthogonal to the CMOs, leading to a Kondo lattice description. 
The resulting quantum criticality has been proposed as the mechanism for 
the strange metal behavior observed in several kagome and pyrochlore metals 
with active flat bands, 
including CuV$_2$S$_4$, Ni$_3$In, Fe$_3$Sn$_2$ and CsCr$_3$Sb$_5$ \cite{Huang2023np,Ye2024,Ekahana2024,Liu2024Superconductivity}. 
A phase diagram that arose from this framework \cite{Chen2023Metallic} 
is consistent with the pressure-induced quantum phase transitions observed in 
a new kagome metal CsCr$_3$Sb$_5$ \cite{Liu2024Superconductivity}. 
These developments, together with the Kondo physics 
suitable for moir\'{e} 
structures \cite{Ram2021,Song2022,Kumar2022,Guerci2023Chiral,Xie-PRR2024,Xie2024Superconductivity,zhao2023gate},
broaden the perspective on the unifying themes for strongly correlated materials and structures \cite{Checkelsky-NRM2024}.

One key distinction of bulk kagome and pyrochlore metals is the existence of a large number 
of bands near the Fermi energy. This complexity makes it difficult to theoretically analyze 
their correlation physics. 
In this work, we show that the CMO enables the understanding of the dominating correlation effect for the low-energy physics. We illustrate our approach in the 
the bilayer kagome metal Ni$_3$In \cite{Ye2024}, in which a large number of bands 
exist near the Fermi energy. Our results provide the basis to
understand the scanning-tunneling-microscopy experiment \cite{exp-stm-2024} 
in this kagome metal in terms of the CMO notion.
The CMO-based description enables us to propose how the strange metallicity in this system
develops based on an overall phase diagram, which
can be tested 
by future experiments.
In the process, we advance a general theoretical approach to understand the active-flat-band physics of bulk materials that is difficult to achieve based on atomic orbitals.

{\it Atomic orbitals, symmetry and star-stacked triangles.~~~}
$\rm Ni_3 In$ has a hexagonal lattice in space group $P6_3/mmc$ (no.~194) \cite{Ye2024}.
Fig.~S1 shows its lattice structure [see the supplemental materials (SM) \cite{supplemental_material}].
A star-stacked triangle describes the
six Ni atoms at $6h$ Wyckoff position,
which are related by the six-fold screw symmetry 
$\{C_{6z}|00\frac12\}$ (six-fold rotation followed by $1/2$-translation along the rotation axis). Its site-symmetry group $mm2$ has four 1d irreducible representations.
Instead of a global basis, it is convenient to work on a local basis such that the local frames of the six sites are also related by the screw symmetry. 
We label the six Ni atoms by $\rm Ni_{j}$, $j=1\dots 6$. 
The $x$-axis of $\rm Ni_{1}$ and $\rm Ni_{2}$ differ by $\pi/3$ rotation.
This choice of the local basis defines the types of orbitals at these sites. For example, the $d_{xz}$ orbitals $|\mathrm{Ni}_{j}^{xz}\rangle$ on the six sites satisfy
$|\mathrm{Ni}_{(j+1 \mod 6)}^{xz}\rangle = \{C_{6z}|00\frac{(-1)^j}{2}\} |\mathrm{Ni}_{j}^{xz}\rangle$.

Each unit cell has $62 = 6\times [1 (\textrm{Ni-}s) + 3 (\textrm{Ni-}p) + 5 (\textrm{Ni-}d)] + 2\times [1 (\textrm{In-}s) + 3 (\textrm{In-}p)]$ valence and conduction orbitals, and $66 = 6\times 10 (\textrm{Ni-}3d^84s^2) + 2 \times3 (\textrm{In-}5s^25p^1)$ valence electrons. 
Based on {\it ab initio} calculation with full-potential local-orbital (FPLO) \cite{Koepernik1999Full,Opahle1999Full,Perdew1996Generalized}, the band structure of $\rm Ni_3 In$ can be well captured by a $62$-orbital tight-binding model, which contains $s$, $p$, $d$ orbitals for each Ni atom, and $s$, $p$ orbitals for each In atom. 
The tight-binding representation shows strong agreement with the ab initio band dispersion, as detailed in the 
SM \cite{supplemental_material}.

\begin{figure}[t]
    \centering
    \includegraphics[width=\linewidth]{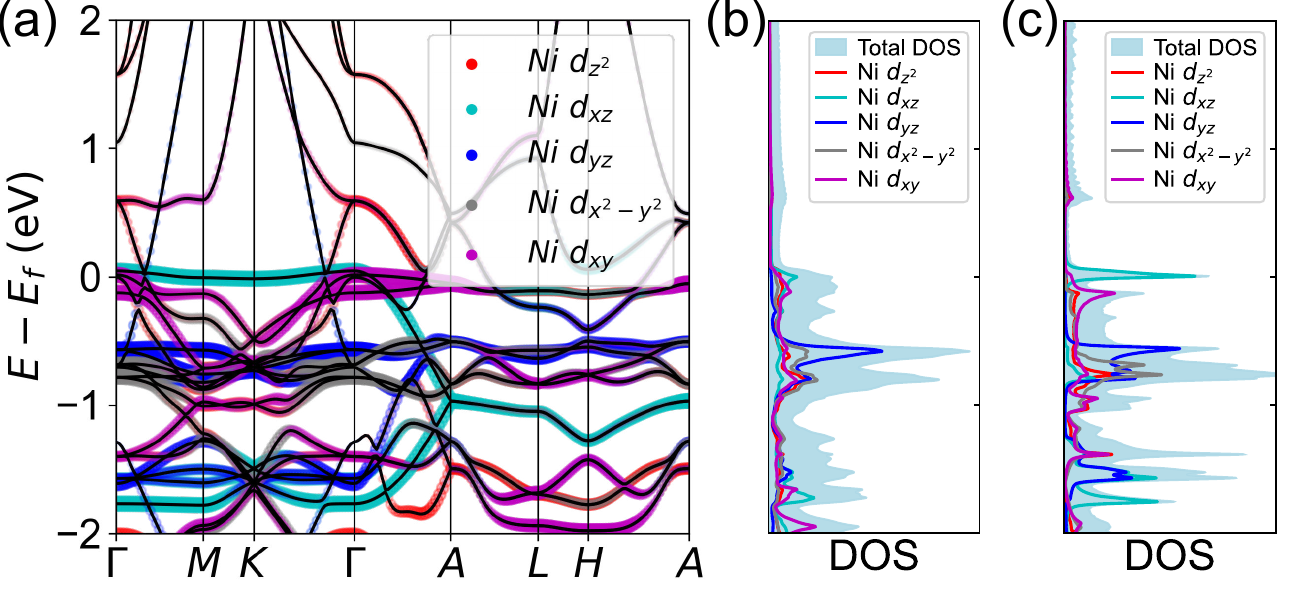}
    \caption{
    (a) {\it Ab initio} band structure of $\rm Ni_3 In$. The color codes represent the orbital content of the $\rm Ni$-$d$ orbitals.
    (b) The density of states of the {\it ab initio} band structure.
    (c) The density of states in the $k_z = 0$ plane. A peak corresponding to Ni-$d_{xz}$ orbitals can be seen at the Fermi level.
    }
    \label{fig:bands}
\end{figure}

Fig.~\ref{fig:bands}(a) shows the the tight-binding representation of the band structure.
The $\rm Ni$ atom $d$-orbital components are represented by different colors, showing that the low energy bands around and below the Fermi energy are mainly composed of the $d$-orbitals. 
The flat band near the Fermi energy (cyan colored band)
predominantly is associated with the 
$d_{xz}$ orbital component.
The 3D density of states (DOS) of the {\it ab initio} band structure, shown in Fig.~\ref{fig:bands}(b), also shows a peak near the Fermi energy.
The $d_{xz}$ flat band is even more pronounced in the 2D DOS of the band structure in the $k_z = 0$ plane, as shown in Fig.~\ref{fig:bands}(c).

{\it Compact molecular orbital.~~~}
The rational of constructing the CMO follows what is prescribed in
Refs.~\cite{Hu2023Coupled, Chen2023Metallic, Chen2024Emergent}. 
With a flat band near the Fermi energy,
we expect the Coulomb interactions to have a dominant effect on 
the real-space states associated with the flat band, but the 
construction of the latter requires that the flat band combines 
with a subset of the wide bands; the CMO corresponds to the ensuing orbital that is dominated by the flat-band states.

The need for such a construction is 
highlighted by 
the large  number of
bands in the tight-binding model; without the anchoring of the flat band, the analysis of the correlation zeffect would be 
very difficult given the large number of involved bands.
We also note that the active flat band lies in the $k_z=0$ plane \cite{Ye2024}
[Fig.~\ref{fig:bands}(a)].
The Hamiltonian of the $k_z=0$ plane is effectively described by the layer group $p6/mmm$.

\begin{figure}[t]
    \centering
    \includegraphics[width=0.75\linewidth]{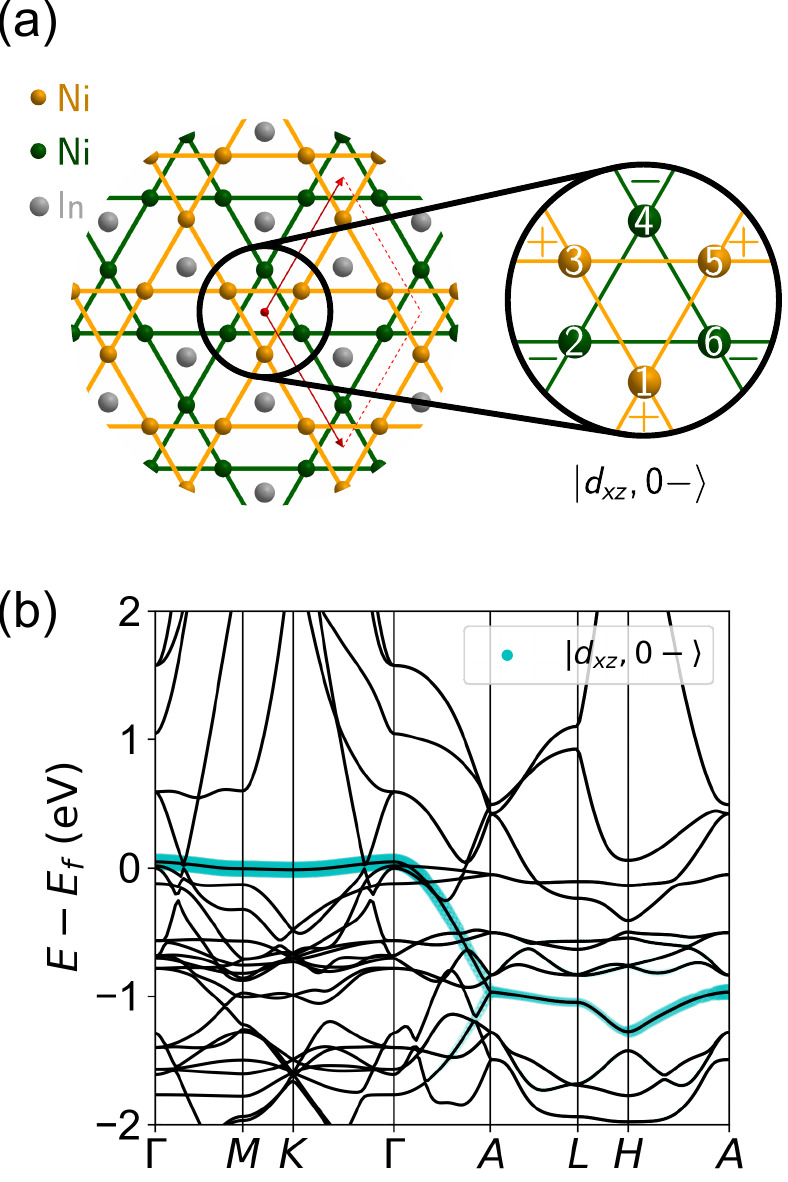}
    \caption{
        (a) The sketch of the wave function $|d_{xz}, 0-\rangle$.
        (b) Band structure projected onto the orbital $|d_{xz}, 0-\rangle$.
    }
    \label{fig:mo}
\end{figure}

To isolate the wide bands that must be taken into account along with the active flat band, so that the dominant correlation effects of the latter can be studied,
we 
start from the elementary band representations (EBRs) that contain the flat band.
As explained in detail in the 
SM \cite{supplemental_material}, 
there is only one EBR that contains the little group representation of the flat band, which is the induced representation $A_{2g} \uparrow G$, and $A_{2g}$ stands for an irreducible representation of the site symmetry group at Wyckoff position $2a$.
The charge center of $2a$ Wyckoff position is located at $(0, 0, 0)$ and $(0, 0, 1/2)$, which are the centers of the two stacked triangles formed by the Ni atoms.
Since there is no atomic orbital at the $2a$ Wyckoff position, 
we expect the CMO wave function to be a linear combination of atomic orbitals 
of the Ni atomic orbitals from the triangles of both layers around the $2a$ 
Wyckoff positions. The $A_{2g}$ symmetry allows the CMO to have components on $d_{xz}$, $d_{xy}$ and $p_x$ orbitals. However, as shown in Fig.~\ref{fig:bands}(a), the flat band at the Fermi energy is mostly formed by the $d_{xz}$ orbital component. 
{In other words, the $d_{xz}$ component of the CMO describes the flat band whose wave function is}:
\begin{align}
    |d_{xz}, 0-\rangle =& \frac{1}{\sqrt{6}}\Big{(}|{\rm Ni}^{xz}_1\rangle + |{\rm Ni}^{xz}_3\rangle + |{\rm Ni}^{xz}_{5}\rangle\nonumber \\
    &- |{\rm Ni}^{xz}_2\rangle - |{\rm Ni}^{xz}_4\rangle - |{\rm Ni}^{xz}_6\rangle \Big{)}\,,\label{eqn:cmo}
\end{align}
in which the notation ``$0-$" indicates the $C_{3z}$ and $C_{2x}$ eigenvalues of the CMO state are $1=e^{i0\times2\pi/3}$ and $-1$, respectively.
The sketch of the wave function $|d_{xz}, 0-\rangle$ is shown in Fig.~\ref{fig:mo}(a).
Additionally, $|d_{xz}, 0-\rangle$ is also constrained in one unit cell, which ensures the orthogonality of the wave functions in different unit cells. 
As a consequence, the CMO in different unit cells are legitimate local orbital basis, and Bloch states can be constructed by Fourier transformation.
Projecting the energy eigenstates obtained from the {\it ab-initio} calculation onto the Bloch state of the CMO orbital reveals a large intensity predominantly in the flat band, as shown in Fig.~\ref{fig:mo}(b).

These properties guarantee that the wave function Eq.~(\ref{eqn:cmo}) provides the most accurate effective description of the flat band degrees of freedom in this material. 
We stress that in our case, the CMO is associated with a star-stacked triangle, which 
is unique to the bilayer kagome system. Importantly, the star-stacked triangles from neighboring unit cells do {\it not} overlap, which ensures that the CMOs from different unit cells are orthogonal to each other. This is qualitatively different from the single layer case.

{\it Anderson lattice model and correlation effects.~~~}
As we already stressed, the basis that contains the CMO state $|d_{xz}, 0-\rangle$ is the most appropriate choice to describe the electronic correlation effects.
To expound on this point, in the atomic basis, the interacting Hamiltonian would contain 
a large amount of orbitals 
and solving for the correlated electronic structure 
is difficult.
In addition to the numerical difficulties, the atomic orbitals  are not the appropriate degrees of freedom that are most relevant to the low-energy physics near the Fermi level.

\begin{figure}[!t]
    \centering
    \includegraphics[width=0.7\linewidth]{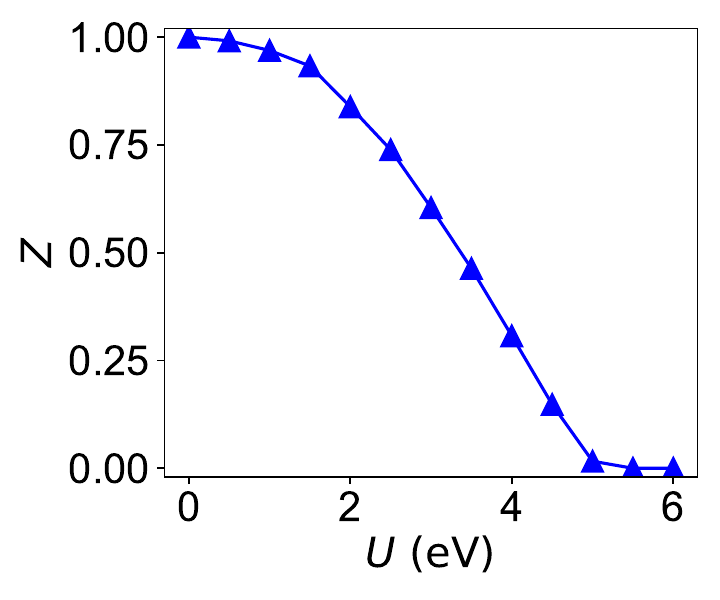}
    \caption{The quasiparticle weight of the CMO state $|d_{xz}, 0-\rangle$ as a function of the Hubbard interaction $U$ on this CMO state.}
    \label{fig:z-vs-U}
\end{figure}

An Anderson lattice model can be constructed with the CMO state treated as the $f$-orbital, which hybridizes with all the other 61 orbitals. 
The advantage of this choice is two-fold.
First, there is only one active orbital per unit cell 
for which the interaction effect is dominantly important.
This leads to drastic 
simplification in the analysis of the strong correlation effect.
Second, the CMO state has a large overlap with the states near the Fermi level as shown in Fig.~\ref{fig:mo}(b), especially in the flat band of the $k_z = 0$ plane; 
this makes it a good choice to describe the low-energy physics.
With this picture in mind, one could rewrite the kinetic Hamiltonian
in the $C_3$ basis as:
\begin{align}
H_0 &= H_f + H_c + H_{fc}  \, , \nonumber \\
H_f &= \sum \limits_{\mathbf{k}} h^{\mathbf{k}}_{ff} f^{\dag}_{\mathbf{k}} f_{\mathbf{k}} \, ,  \\
H_c &= \sum \limits_{\mathbf{k}} \sum \limits_{\alpha,\beta \ne f} h^{\mathbf{k}}_{\alpha\beta} c^{\dag}_{\mathbf{k}\alpha} c_{\mathbf{k}\beta} \, ,  \\
H_{fc} &= \sum \limits_{\mathbf{k}} \sum \limits_{\alpha \ne f} \left( h^{\textbf{k}} _{\alpha f} c^{\dag}_{\textbf{k}\alpha} f_{\textbf{k}} + h^{\textbf{k}} _{f\alpha} f^{\dag} _{\textbf{k}} c_{\textbf{k}\alpha} \right) \,  .
\end{align}
Here, $H_f$ contains the hopping terms among the $f$ orbitals, $H_{fc}$
those between the $f$ and other orbitals, and $H_c$
all the terms that 
do not involve the $f$ orbital 
[written in the notation of Supplemental Materials Eq.~(S1)]. 
{For any $\mathbf{k}$, the conduction bath hamiltonian $h^{\mathbf{k}}_{\alpha\beta}$ ($\alpha, \beta \ne f$)is a $61\times 61$ matrix while the hybridization $h^{\mathbf{k}}_{\alpha f}$ is a $61\times 1$ column matrix.} A Hubbard $U$ term 
is retained for the CMO, which can be written as:
\begin{equation}
    H_I = U \sum_{i} \tilde{n}_{fi\uparrow}\tilde{n}_{fi \downarrow}\,,
\end{equation}
in which $\tilde{n}_{f i \sigma} = f^\dagger_{i \sigma} f_{i\sigma} - \frac12$ is the relative density operator of the corresponding CMO state in unit cell $i$.
We note that the value of $U$ is different from the on-site interaction of the atomic $d$ orbitals, since the CMO wave function spreads over multiple atomic sites.

The interacting Hamiltonian can be solved using the $U(1)$ slave-spin approach \cite{Yu2012U1slave}.
In the framework of slave-spin, the quasiparticle weight and 
on-site potential energy shift can be obtained for 
the $f$ orbital in a self-consistent manner.
The correlation effect 
determines a quasiparticle weight smaller than unity, which indicates the suppression of the strength of electron hoppings in the $f$ orbital, and the coherent motion of the electrons.

Here, we solved the interacting Hamiltonian with interaction strength $0 \leq U \leq 6\,\rm eV$.
The quasiparticle weight of the CMO as a function of $U$ is shown in Fig.~\ref{fig:z-vs-U}.
It systematically decreases with the increase of $U$, and it almost vanishes at $U \approx 5\,\rm eV$.
This result provides a quantitative understanding of the correlation effect in this material and demonstrates that the CMO state is indeed the proper degrees of freedom to describe the relevant correlation physics.
For reasonable values of the Coulomb interaction, our calculation suggests a correlation-induced band narrowing that is compatible with the corresponding factor determined by 
thermodynamic \cite{Ye2024} and STM measurements \cite{exp-stm-2024}.

\begin{figure}[!t]
\centering
\includegraphics[width=\linewidth]{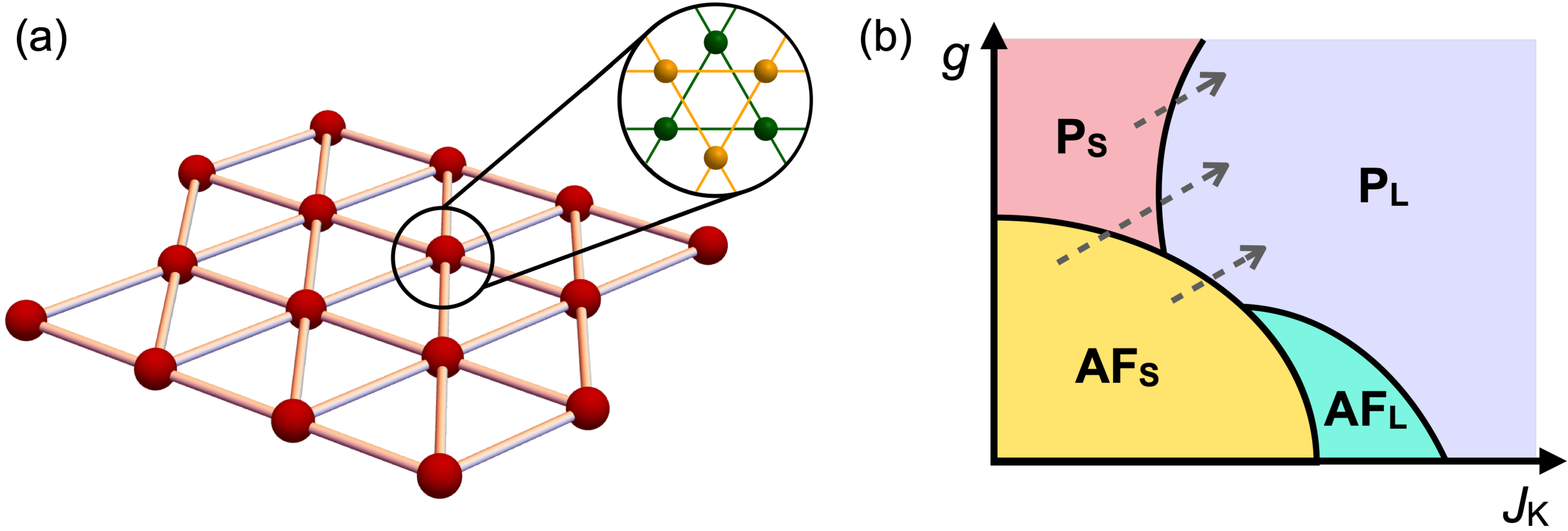}
\caption{Effective model in terms of the compact molecular orbitals (CMOs) and phase diagram. (a) The CMOs (inset; cf. Fig.~\ref{fig:mo}), localized on each plaquette of star stacked triangles of the kagome-bilayer in Ni$_3$In, form an effective triangular lattice. (b) The phase diagram  in the presence of frustrated antiferromagnetic interactions ($g$ parameterized the extent of frustration) and the Kondo-hybridization interaction ($J_K$). Here, `AF$_\text{S}$' (`AF$_\text{L}$') refers to a metallic antiferromagnet with a small (large) Fermi surface, for which the local moments do not (do) participate in the Fermi volume, and `P$_\text{S}$' (`P$_\text{L}$') 
denote the metallic paramagnet counterparts. 
The dashed lines describe the three possible trajectories of quantum phase transitions  proposed for Ni$_3$In.}
\label{fig:triangular}
\end{figure}

{\it Implication for the phase diagram.~~~}
The CMOs we have constructed 
form an effective triangular lattice as depicted in Fig.~\ref{fig:triangular}(a). 
Thus, the quantum magnetism of the local moments features geometrical frustration.
For Kondo lattice 
systems, the effect of frustration is 
summarized in the global phase diagram, Fig.~\ref{fig:triangular}(b), as previously considered theoretically \cite{paschen_quantum_2021,Si-physicab-06,Si_PSSB10,Coleman_Nev,Pixley-2014}
and  explored experimentally \cite{Zhao2019Quantum,Custers2012Destruction,Custers2010,friedemann2009detaching}.
Here $g$ parametrizes 
the degree of frustration and $J_K$ is the Kondo-hybridization coupling. 
With adequate frustration, we can expect the quantum phases and their transitions to occupy the upper portion of the phase 
diagram. The quantum phase transition, expected from the competition between the Kondo and RKKY interactions that involve the CMOs, is then 
expected to follow
three possible trajectories as a function of external control parameters, as shown in Fig.~\ref{fig:triangular}(b). 
Accordingly, strange metallicity can develop either 
from an underlying quantum critical point or a quantum
critical phase. Which of these routes operate in Ni$_3$In
can be elucidated from
tuning Ni$_3$In by such means as tensile stress or chemical substitution.

{\it Discussion.~~~}
Several remarks are in order, First,
our work represents the first materials-based construction of the CMO state.
Our case study is representative of bulk frustrated-lattice metals.
Namely, 
there is an abundance of bands near the Fermi energy, which makes it difficult to 
study correlation physics using atomic orbitals.
Our work shows that, through the CMO construction, 
a flat band enables the resolution of this challenge.
Because the flat band captures the dominating correlation effect,
it provides a means to select 
the wide bands that are important for the correlated low-energy physics.
These are the bands that intermix with the flat band the most, 
and they can be determined based on the consideration of symmetry and topology.

Second,
our materials-based construction of the CMO state has clear-cut implications for experiments.
Since the flat band CMO state $|d_{xz}, 0-\rangle$ involves atomic orbitals 
from different sites of the unit cell with definite a phase relationship,
we can expect that spectroscopy measurements with atomic resolutions will
observe sharply contrasting spectra at the different sites of the unit cell.
Indeed, our theoretical description 
provides the basis to understand 
this type of STM experiments in Ni$_3$In, as described elsewhere
\cite{exp-stm-2024}. 

{\it Summary.~~~}
In conclusion, we have proposed a compact molecular orbital that captures the flat band near the Fermi level in $\rm Ni_3 In$, 
recognizing that the bilayer nature of the lattice structure simplifies its construction.
This CMO state is a linear combination of the $d_{xz}$ orbitals on the six Ni atoms, whose charge center is not at any atomic states. 
We further utilized this CMO state to construct an Anderson lattice model, which enables a theoretical analysis of the Coulomb interaction effects in this system that would have been much more difficult in terms the atomic orbitals. The implications of our results are multi-fold. From a pragmatic perspective, our explicit construction of the CMOs allows for their detection by atomic-resolution experiments \cite{exp-stm-2024}. Theoretically,
 our work
demonstrates the first materials-specific setting that showcases how the CMOs 
can be used to enable the theoretical description
of the correlation effects in realistic frustrated-lattice metals that often involve many bands near the Fermi energy. More generally, our work illustrates how local degrees of freedom can emerge in very complex quantum materials that enable the understanding of their quantum phase transitions and strange metallicity on par with what happens in a variety of more established systems of strong correlations. Accordingly,
our work raises the prospect to realize a universal understanding of
physics in the regime of amplified quantum fluctuations across the strongly correlated electron systems.

\begin{acknowledgments}
We thank Gabriel Aeppli, Jennifer Cano, Pengcheng Dai, Ming Yi and, 
especially, Roser Valent\'\i, 
for useful discussions.
Work at Rice has primarily been supported by the U.S. DOE, BES, under Award No. DE-SC0018197 
(model construction, M.M. and F.X.),
by the Air Force Office of Scientific Research under Grant No. FA9550-21-1-0356 (electronic structure analysis, Y.F.),
by the Robert A. Welch Foundation Grant No. C-1411 (correlation calculation, M.M. and L.C.), and 
by the Vannevar Bush Faculty Fellowship ONR-VB N00014-23-1-2870 (model conceptualization, Q.S.).
The majority of the computational calculations have been performed on the Shared University Grid
at Rice funded by NSF under Grant EIA-0216467, a partnership between Rice University, Sun
Microsystems, and Sigma Solutions, Inc., the Big-Data Private-Cloud Research Cyberinfrastructure
MRI-award funded by NSF under Grant No. CNS-1338099, and the Extreme Science and
Engineering Discovery Environment (XSEDE) by NSF under Grant No. DMR170109. 
The work in Vienna was supported by the Austrian Science Fund (FWF grants SFB F86 ``Q-M\&S'', I5868-N/FOR5249 ``QUAST'', and 10.55776/COE1 ``quantA'') and the European Research Council (ERC Advanced Grant 101055088-CorMeTop). {HB, NA, and JCS acknowledge funding by the BSF-NSF-Materials grant number 2020744. JCS acknowledges support from the Paulo Pinheiro de Andrade fellowship.}
Q.S. acknowledge the hospitality of the Aspen Center for Physics, which is supported by NSF grant No. PHY-2210452.
\end{acknowledgments}

\bibliography{references.bib}
\bibliographystyle{apsrev4-2}

\onecolumngrid
\newpage
\beginsupplement

\section*{Supplemental Materials}

\section{Lattice structure}
The lattice structure of $\rm Ni_3 In$
is shown in Fig.~1(a).
The six $\rm Ni$ atoms are located at the $6h$ Wyckoff position 
(of multiplicity $6$).
in the upper layer with coordinate $z=3/4$ and the remaining three are in the lower layer at $z=1/4$. The yellow and green symbols represent the $\rm Ni$ atoms in the upper and lower layers, respectively. The $\rm Ni$ atoms in each layer form a breathing kagome lattice. The two layers are related by the six-fold screw symmetry.
The  two $\rm In$ atoms sit at the $2c$ Wyckoff position (of multiplicity $2$), 
which are represented by gray symbols. 
One of the sites is in the upper layer
and the
other is in the lower layer.
They are at the kagome center, forming a triangular lattice.

\begin{figure}[h]
    \centering
    \includegraphics[width=0.5\linewidth]{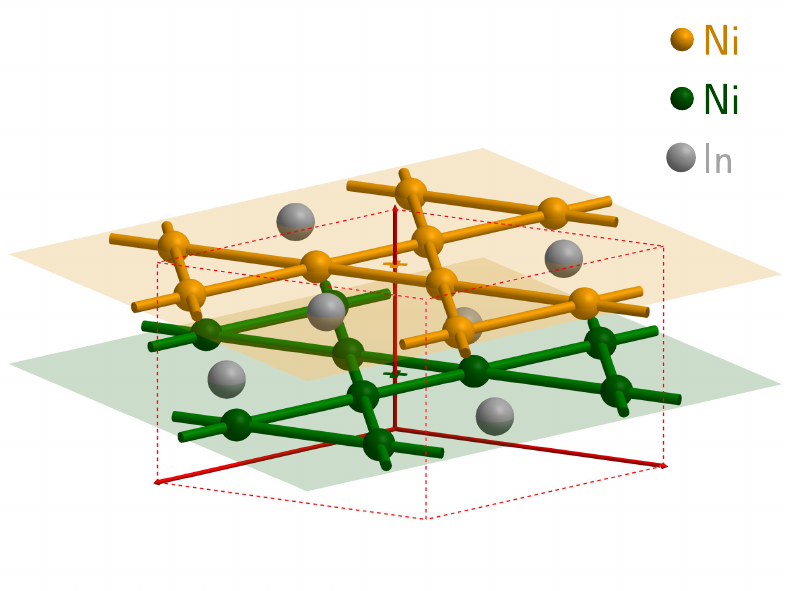}
    \caption{The lattice structure of Ni$_3$In. Yellow and green spheres represent Ni atoms, while grey spheres represent In atoms. The red arrows stand for the basis vectors of the Bravais lattice.}
    \label{fig:lattice}
\end{figure}

\section{Tight-binding model} \label{supp:tight-binding}

We consider a tight binding model made of $\mathcal{N}$ unit cells, each containing 6 Ni and 2 In atoms. The model has 62 orbitals by considering $s$, $p$, $d$ orbitals of each Ni atom, and $s$, $p$ orbitals of each In atom. Let $|\varphi_{i\alpha}\rangle$ be the wavefunction of orbital $\alpha$ at unit cell $\mathbf{R}_i$. Within the unit cell at $\mathbf{R}_i$, the orbital $|\varphi_{i\alpha}\rangle$ is located at $\boldsymbol{\tau}_{\alpha}$. The hamiltonian is:
\begin{equation}
    H_0 = \sum\limits_{i\alpha, j\beta} t_{\alpha\beta}(\mathbf{R}_i - \mathbf{R}_j) c^{\dag}_{i\alpha}c_{j\beta} = \sum\limits_{\mathbf{k}} \sum\limits_{\alpha\beta} h^{\mathbf{k}}_{\alpha\beta} c^{\dag}_{\mathbf{k}\alpha} c_{\mathbf{k}\beta} \,, \label{eqn:non-interacting-hamiltonian} 
\end{equation}
where $c^{\dag}_{\mathbf{k}\alpha} = \frac{1}{\sqrt{\mathcal{N}}} \sum \limits_i e^{i \mathbf{k} \cdot \left( \mathbf{R}_i + \boldsymbol{\tau}_{\alpha} \right)} c^{\dag}_{i\alpha}$ is the creation operator for the $\textbf{k}$ momentum mode, and
\begin{equation}
      h^{\mathbf{k}}_{\alpha\beta} = \sum\limits_{\mathbf{R}}  t_{\alpha\beta}(\mathbf{R}) e^{- i \mathbf{k} \cdot \left( \mathbf{R} + \boldsymbol{\tau}_{\alpha} - \boldsymbol{\tau}_\beta \right) } \,.
    \end{equation}
For each $\mathbf{k}$, the matrix $h^{\textbf{k}}$ is a $62\times 62$ matrix with 62 eigenvalues $\left\{E_{n\textbf{k}}\right\}_{n=1}^{62}$, and corresponding eigenstates $\left\{|\psi_{n\textbf{k}}\rangle\right\}_{n=1}^{62}$.

\begin{figure}[h]
    \centering
    \includegraphics[width=0.7\linewidth]{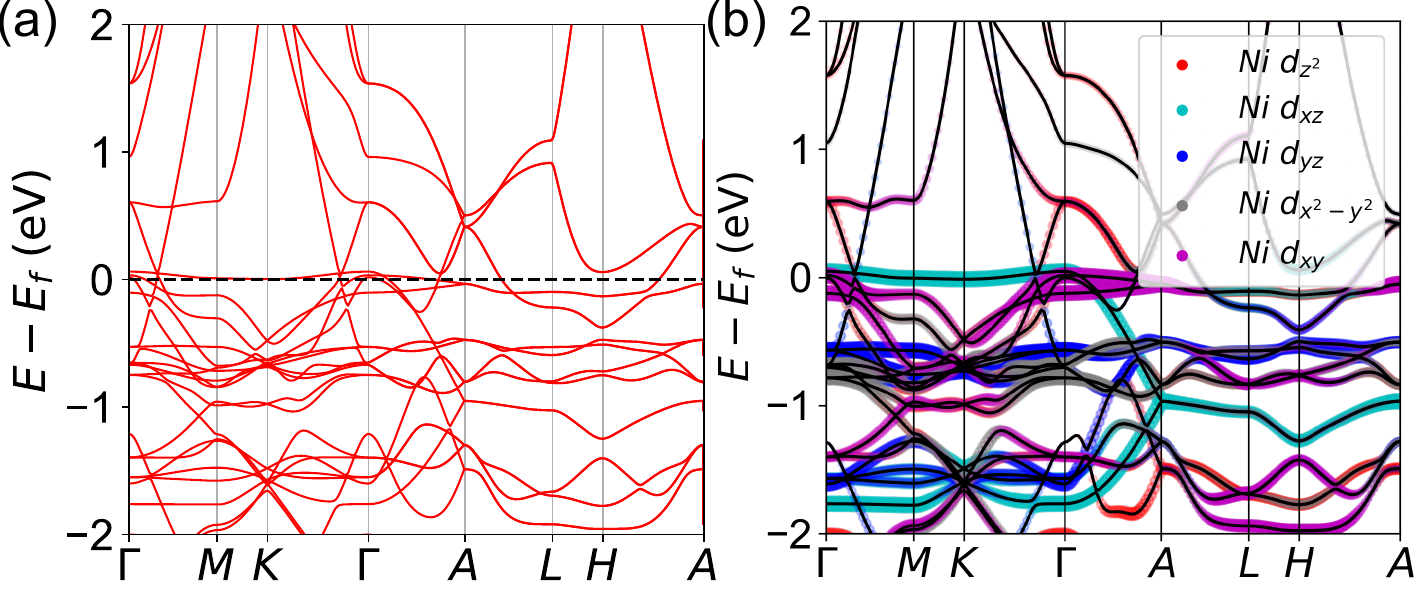}
    \caption{(a) The {\it ab initio} band structure of Ni$_3$In, computed by full-potential local-orbital method \cite{Koepernik1999Full,Opahle1999Full,Perdew1996Generalized}. 
    (b) The band structure of the $62$-orbital tight-binding model. The $d$ orbitals from the Ni atoms are labeled with different colors.
    }
    \label{fig:dft}
\end{figure}

\section{Elementary band representation}

The goal is to look for Elementary Band Representations (EBRs) that contain the flat band near the fermi level. The $\Gamma$, $M$, $K$ irreps of the flat band are $\Gamma_{4+}$, $M_{4+}$, $K_3$ respectively. There are 2 indecomposable EBRs that contain these irreps: $A_{2g} \uparrow G$, $A_2^{\prime} \uparrow G$ (summarized in Table~\ref{tab:ebr-irreps}). Of the two, $A_{2g}^{\prime}$ is not possible as no eigenstate at $\Gamma$ point is a $\Gamma_{1-}$ irrep. This leaves $A_{2g} \uparrow G$ as the smallest EBR to contain the flat band.

\begin{table}[h]
    \begin{tabular}{l|cc}
    \hline
    & $A_{2g}$ & $A_2^{\prime}$\\
    \hline
    $\Gamma$    & $\Gamma_{2+} \oplus \Gamma_{4+}$ & $\Gamma_{1-} \oplus \Gamma_{4+}$\\
    $M$    & $M_{2+} \oplus M_{4+}$ & $M_{1-} \oplus M_{4+}$ \\
    $K$   & $K_3 \oplus K_4$ & $K_2 \oplus K_3$ \\
    $A$    & $A_2$ & $A_2$\\
    $L$  & $L_2$ & $L_2$\\
    $H$     & $H_3$ & $H_3$ \\
    \hline
    \end{tabular}
\caption{\label{tab:ebr-irreps} Elementary band representations that contain all the irreps of the flat band (adapted from Bilbao crystallographic server \cite{aroyoBilbaoCrystallographicServer2006a, aroyoBilbaoCrystallographicServer2006})}
\end{table}

\begin{figure}[t] 
\includegraphics[width=0.5\linewidth]{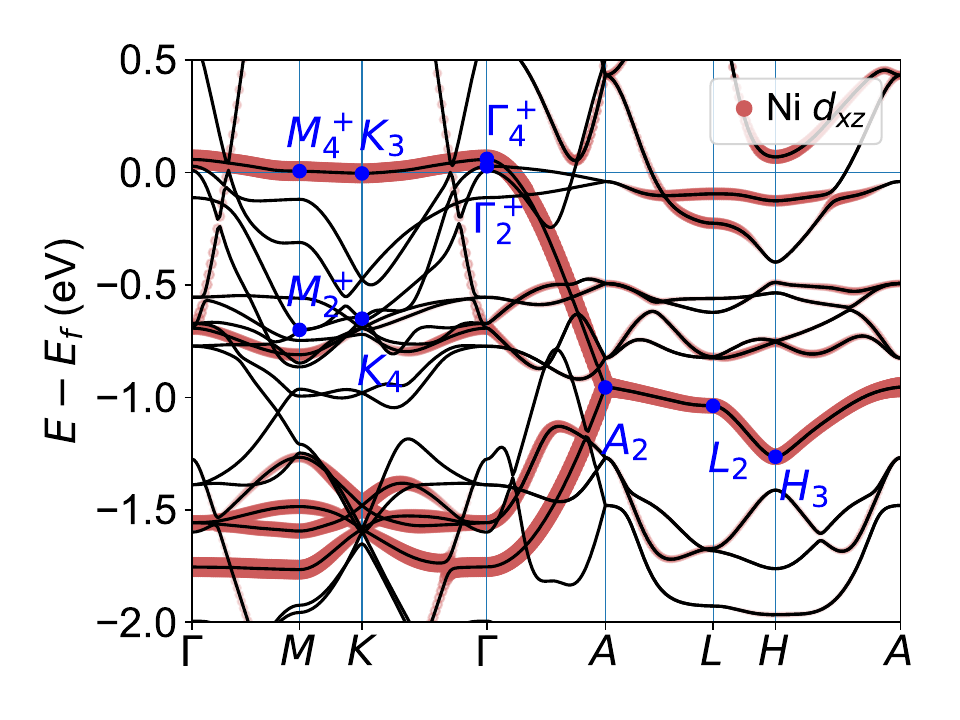}
 \caption{Band structure of Ni$_3$In with the band structure of Ni $d_{xz}$ orbitals in the local basis. The blue points are the irreps considered in constructing the EBR capturing the flat band.}
    \label{fig:enter-label}
\end{figure}
Band representation $A_{2g} \uparrow G$ has a site symmetry group $\bar{3}m$ with $2a$ wyckoff position located at  $\boldsymbol{\tau}_1 = \left( 0,0,0 \right) $, $\boldsymbol{\tau}_2 = \left( 0,0, 1/2 \right) $ (in the $\left\{ \mathbf{a}_i \right\}_{i=1}^3$ basis). Let $|\phi^1_{A_{2g}}\rangle$, $|\tilde{\phi}_2\rangle$ be the molecular orbitals described by $A_{2g}$ at $\boldsymbol{\tau}_1$, $\boldsymbol{\tau}_2$ respectively. The generators of site symmetry group $\bar{3}m$ at $\boldsymbol{\tau}_1$ are $\left\{ C_{3z}, C_{2x}, C_{2y}, C_{2,x+y}, I \right\}$. Orbital $|\phi^1_{A_{2g}}\rangle$ is a linear combination of atomic orbitals such that $\langle\tilde{\phi}_1| \mathbf{\hat{r}} |\phi^1_{A_{2g}}\rangle = \boldsymbol{\tau}_1$  and
\begin{align}
  &C_{3z} |\phi^1_{A_{2g}}\rangle = |\phi^1_{A_{2g}}\rangle, \quad I |\phi^1_{A_{2g}}\rangle = |\phi^1_{A_{2g}}\rangle\,, \\
    &C_{2x} |\phi^1_{A_{2g}}\rangle = -|\phi^1_{A_{2g}}\rangle, \quad
    C_{2y} |\phi^1_{A_{2g}}\rangle = -|\phi^1_{A_{2g}}\rangle, \quad C_{2, x+y} |\phi^1_{A_{2g}}\rangle = -|\phi^1_{A_{2g}}\rangle
\end{align}
Under these constraints, the minimal form of $|\phi^1_{A_{2g}}\rangle$ consists of orbitals from the nearest atoms to $\boldsymbol{\tau}_1$, i.e., 6 Ni atoms around $(0, 0, 0)$, given by 
\begin{align}
    |\phi^1_{A_{2g}} \rangle =& \frac{A_{xz}}{\sqrt{6}} \Big( |\mathrm{Ni}_1^{xz}\rangle + |\mathrm{Ni}_3^{xz}\rangle + |\mathrm{Ni}_5^{xz}\rangle - |\mathrm{Ni}_2^{xz}\rangle -  |\mathrm{Ni}_4^{xz}\rangle - |\mathrm{Ni}_6^{xz}\rangle \Big) \nonumber\\
    & + \frac{A_{xy}}{\sqrt{6}} \Big( |\mathrm{Ni}_{1}^{xy}\rangle + |\mathrm{Ni}_{3}^{xy}\rangle + |\mathrm{Ni}_5^{xy}\rangle + |\mathrm{Ni}_{2}^{xy}\rangle + |\mathrm{Ni}_4^{xy}\rangle + |\mathrm{Ni}_6^{xy}\rangle \Big)  \nonumber\\
    & + \frac{A_{px}}{\sqrt{6}} \Big( |\mathrm{Ni}_1^{px}\rangle + |\mathrm{Ni}_3^{px}\rangle + |\mathrm{Ni}_5^{px}\rangle  + |\mathrm{Ni}_2^{px}\rangle + |\mathrm{Ni}_4^{px}\rangle + |\mathrm{Ni}_6^{px}\rangle \Big)\,. \label{eqn:cmo-general}
\end{align}
Normalizing $|\phi^1_{A_{2g}}\rangle$ means $\left| A_{xz} \right|^2 + \left| A_{yz} \right|^2 + \left| A_{px} \right|^2 = 1$. Note that in constructing the molecular orbital, we considered the closest atoms to the $2a$ wyckoff position, i.e., 6 Ni atoms as two stacked triangles. However, one could also consider the next nearest neighbours and so on giving a tail like terms to $|\phi^1_{A_{2g}}\rangle$ in Eq.~(\ref{eqn:cmo-general}). Fig.~\ref{fig:A2g-orb-comp} shows different band structure of different orbital components of $|\phi^1_{A_{2g}}\rangle$ clearly indicating the $|d_{xz}, 0-\rangle$ captures the flat band.

\section{Using $C_3$ basis}

\begin{figure*}[htbp] \centering
\includegraphics[width=0.7\linewidth]{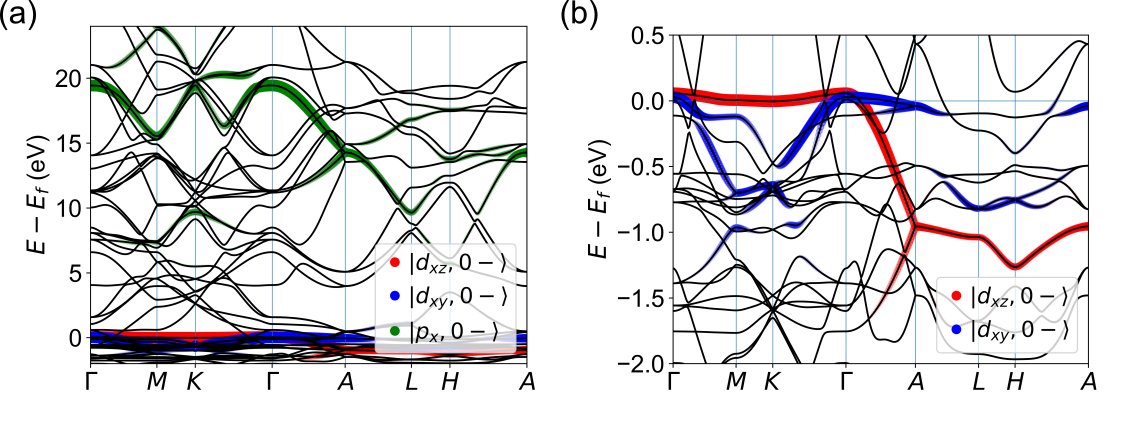}
\caption{Band structure of different components of $|\phi^1_{A_{2g}}\rangle$ (a) $|p_x, 0-\rangle$, $|d_{xz}, 0-\rangle$, $|d_{xy}, 0-\rangle$ (b) $|d_{xz}, 0-\rangle$, $|d_{xy}, 0-\rangle$ near the Fermi level showing how $|d_{xz}, 0-\rangle$ captures the flat band. \label{fig:A2g-orb-comp}}
\end{figure*}

\begin{figure} \centering
\includegraphics[width=0.7\linewidth]{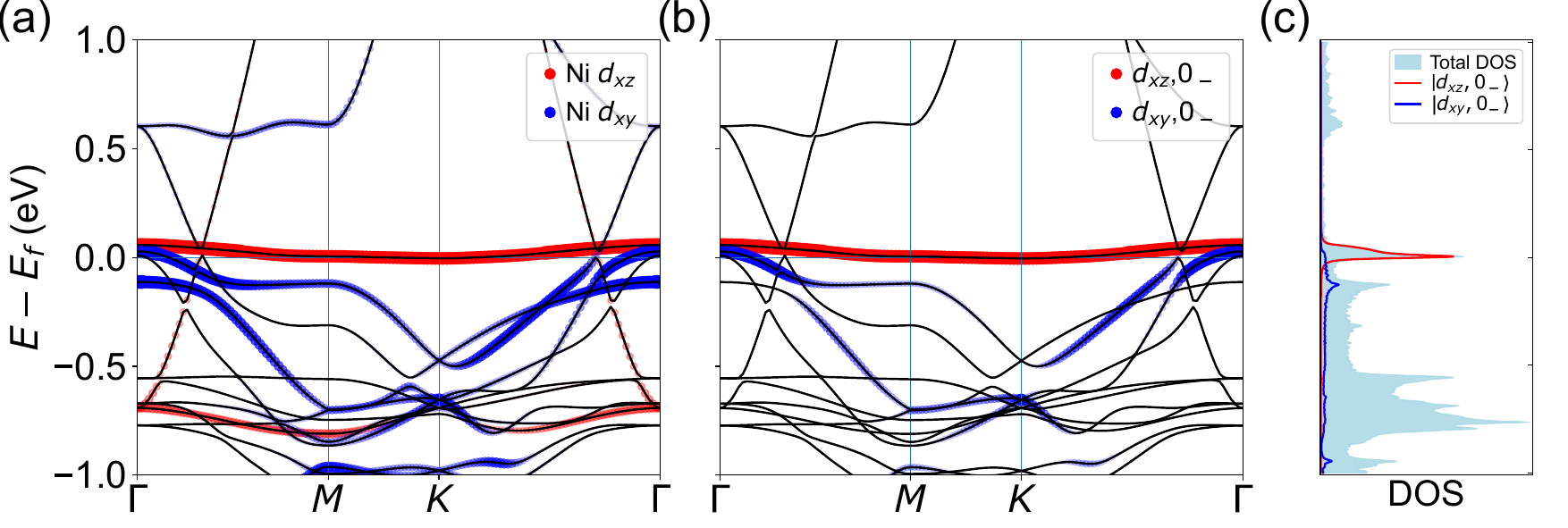}
\caption{Band structure of $k_z=0$ plane with orbital content of (a) Ni $d_{xz}$, $d_{xy}$ orbitals in the atomic orbital basis (b) $| d_{xz}, 0_- \rangle$, $|d_{xy}, 0_-\rangle$ molecular orbitals. (c) Density of states in the $k_z=0$ plane (normalized to 1) with orbital content of molecular orbitals  $| d_{xz}, 0_- \rangle$, $|d_{xy}, 0_-\rangle$.}
    \label{fig:dos}
\end{figure}

To examine the behaviour of tunneling into the molecular orbitals, we choose a unitary transformation such that different orbital components of $|\phi^1_{A_{2g}}\rangle$ are part of the basis after the transformation. One such choice are the eigenbasis of $C_{3z}$ being one of the generators of site symmetry group at $\boldsymbol{\tau}_1$. The matrix form of $C_{3z}$ on one orbital species of Ni atoms is given by
\begin{equation*}
  C_{3z} = \begin{pmatrix}
      0 & 1 & 0 &  &  &  \\
      0 & 0 & 1 &  &  &  \\
      1 & 0 & 0 &  &  &  \\
      & & & 0 & 0 & 1 \\
      & & & 1 & 0 & 0 \\
      & & & 0 & 1 & 0 \\ \end{pmatrix}
  \end{equation*}
where the basis order is $\big\{$Ni$_1$,  Ni$_3$, Ni$_5$, Ni$_2$, Ni$_4$, Ni$_6$$\big\}$ which are the Ni atoms making up the green, pink triangles around $\boldsymbol{\tau}_1$. The eigenstates, eigenvalues of $C_{3z}$ are
\begin{equation*}
C_{3z} |m\rangle = e^{i \frac{2\pi m}{3}} |m\rangle \implies |m\rangle = 
\frac{1}{\sqrt{3}}  \begin{pmatrix} a \\ ae^{i \frac{2 \pi m}{3}} \\ ae^{i \frac{4 \pi m}{3}} \\ b \\ be^{i \frac{4 \pi m}{3}} \\ b e^{i \frac{2\pi m}{3}} \end{pmatrix}
\end{equation*}
where $a, b \in \mathds{C}$ and $\left| a \right|^2 + \left| b \right|^2 = 1$. Note that eigenspace of each eigenvalue is 2-dimensional. Considering the form of $|\phi^1_{A_{2g}}\rangle$, we choose $a = \pm b = \frac{1}{\sqrt{2}}$ for the new basis of Ni orbitals. Each state in this basis is labelled by $|l, m \pm \rangle$ where $l \in \left\{ s, p_x, p_y, p_z, d_{xz}, d_{yz}, d_{xy}, d_{x^2-y^2}, d_{z^2} \right\}$ of Ni atoms around $\boldsymbol{\tau}_1$ in the local basis, $m$ is the eigenvalue of $C_{3z}$, sign $\pm$ is assigned based on its eigenvalue under $C_{2x}$. The orbital $|\phi^1_{A_{2g}}\rangle$ (described in Eq.~\eqref{eqn:cmo-general}) in this basis is
\begin{equation}
  |\phi^1_{A_{2g}} \rangle = A_{xz} |d_{xz}, 0-\rangle  + A_{xy} |d_{xy}, 0-\rangle + A_{px} |p_x, 0-\rangle    \label{eq:408} 
\end{equation}
Fig.~\ref{fig:A2g-orb-comp} shows the band structure of  Ni $d_{xz}$, $d_{xy}$, $p_x$ components of $|\phi^1_{A_{2g}}\rangle$ demonstrating how $d_{xz}$, $d_{xy}$ orbitals are close to the fermi level while $p_x$ has very high energy. Fig.~\ref{fig:A2g-orb-comp} shows how $d_{xz}$ component of $|\phi^1_{A_{2g}}\rangle$ captures the flat band completely further confirmed $k_z=0$ plane density of states show in Fig.~\ref{fig:dos}.

%

\section{Slave-spin approach}
In this section, we provide a brief review of the $U(1)$ slave spin approach, which is widely used in studying strongly correlated electronic systems.

We consider a multi-orbital interacting model with Hamiltonian 
\begin{equation}\label{eqn:FullHam}
    H = H_0 + H_1\,.
\end{equation}
Here the tight-binding Hamiltonian $H_0$ is given by Eq.~(\ref{eqn:non-interacting-hamiltonian}), and the interaction Hamiltonian is given by:
\begin{align}\label{eqn:interaction}
    H_I = \frac{U}{2} \sum_{i,\sigma} \tilde{n}_{if\sigma} \tilde{n}_{if\bar{\sigma}} \,,
\end{align}
where $\tilde{n}_{if\sigma}=f^\dagger_{i\sigma}f_{i\sigma} - \frac12$ is the fermion density of the molecular orbital and $U$ is the Hubbard interaction strength.

In the $U(1)$ slave spin approach, the local fermionic operator $f^\dagger_{j\alpha\sigma}$ is mapped to a product of two parton operators:
\begin{equation}\label{eqn:parton}
    f^\dagger_{j0\sigma} \rightarrow \eta^\dagger_{j\sigma}o^\dagger_{j\sigma}\,.
\end{equation}
Here, $\eta^\dagger_{j\sigma}$ is fermionic, $o^\dagger_{j\sigma}$ is a spin-$\frac12$ operator and the spin operator $o^\dagger$ has the following form:
\begin{equation}\label{eqn:slave_o}
    o^\dagger_{j\sigma} = P^+_{j\sigma} S^+_{j\sigma}P^-_{j\sigma}\,,~~~P^\pm_{j\sigma} = \frac{1}{\sqrt{\frac{1}{2} \pm S^z_{j\sigma}}}\,.
\end{equation}
To preserve the local physical Hilbert space, we introduce Lagrange multipliers terms for each orbital into the parton Hamiltonian:
\begin{equation}\label{eqn:lagrange_mult}
    H_\lambda = \sum_{j\sigma} \lambda\left(S^z_{j\sigma} + \frac12 - \eta^\dagger_{j\sigma} \eta_{j\sigma}\right)\,,
\end{equation}
and the local constraint $\langle S^z_{j\sigma} \rangle + 1/2 = \langle \eta^\dagger_{j\sigma} \eta_{j\sigma} \rangle$ can be satisfied on the mean field level by controlling the values of $\lambda$. 
Plug Eqs.~(\ref{eqn:parton}-\ref{eqn:lagrange_mult}) into Eqs.~(\ref{eqn:FullHam}-\ref{eqn:interaction}), we get the parton (slave-spin) representation of the interacting Hamiltonian. Then we solve the ground state of Eq.~(\ref{eqn:FullHam}) in this representation.

Using “mean-field” assumption and a “single-site approximation” for the parton operators, we have the following effective decoupled mean-field Hamiltonians for slave-spins and slave-fermions:
\begin{align}
    H^S &= U\sum_{\alpha}S^z_{\alpha, \uparrow} S^z_{\alpha\downarrow} 
    + \sum_{\sigma}\lambda_\alpha S^z_{\sigma} + \sum_{ \sigma}\left[h_0 \frac{S^+_{\sigma}}{\sqrt{n_0(1 - n_0)}} + {\rm H.c.}\right]\,, \\
    H^f &= \sum_{\vk\alpha\beta\sigma}h^f_{\alpha\beta}(\vk) \eta^\dagger_{\vk\alpha\sigma}\eta_{\vk \beta\sigma}\,,
\end{align}
where the bath field $h_\alpha$ and the slave-fermion Hamiltonian $h^f_{\alpha\beta}(\vk)$ are given by
\begin{align}
    h_\alpha &= \frac{1}{N}\sum_{\mathbf{k}\beta}t_{\alpha\beta}(\vk)\sqrt{Z_\beta} \langle \eta^\dagger_{\vk\alpha\sigma} \eta_{\vk\beta\sigma} \rangle\,, \\
    h^f_{\alpha\beta}(\vk) &= \sqrt{Z_\alpha Z_\beta}t_{\alpha\beta}(\vk) + \delta_{\alpha\beta}(\tilde{\varepsilon}_\alpha - \lambda_\alpha + \lambda^0_\alpha - E_F)\,, \\
    \lambda^0_\alpha &= -\sqrt{Z_\alpha}|h_\alpha|\frac{2 n_\alpha - 1}{n_\alpha(1 - n_\alpha)}\,, \\
\end{align}
and the quasiparticle weight $Z_\alpha$ is determined by
\begin{align}
    &\sqrt{Z_{\alpha}} = \langle o^\dagger_{i\alpha\sigma} \rangle = \langle o^\dagger_{i\alpha \bar{\sigma}}\rangle = \langle o^\dagger_{j\alpha\sigma} \rangle \,, \\
    &o^\dagger_{j\alpha\sigma} \approx \frac{S^+_{j\alpha\sigma}}{\sqrt{n_\alpha(1 - n_\alpha)}} + \sqrt{Z_{\alpha}}\eta_\alpha\left[2S^z_{j\alpha\sigma} - (2n_\alpha - 1) \right],\label{eqn:mf-decoupling}
\end{align}
in which $\eta_\alpha = \frac12 \frac{n_\alpha - 1/2}{n_\alpha(1 - n_\alpha)}$ and $n_\alpha = \langle \eta^\dagger_{j\alpha\sigma} \eta_{j\alpha\sigma} \rangle$ is the fermion density.

For a given total filling factor, the mean-field parameters $\sqrt{Z_\alpha}$, $\lambda_\alpha$ and $\lambda^0_\alpha$ can be solved self-consistently.
Then other parameters $n_\alpha$, $E_F$, $h_\alpha$, $\epsilon_\alpha$, etc. can all be obtained.



\end{document}